\documentclass[12pt,a4paper]{cibb}

\usepackage{subfigure,graphicx}
\RequirePackage{multirow}
\usepackage{amsmath,amsfonts,latexsym,amssymb,euscript,xr}
\usepackage{breakcites}

\RequirePackage[colorlinks,citecolor=blue,urlcolor=blue]{hyperref}

\usepackage{xcolor}

\newcommand{\mo}[1]{{#1}}

\title{\large $\ $\\ \bf Composite local low-rank structure in learning drug sensitivity}

\author{The Tien Mai$^{(1)}$, Leiv R{\o}nneberg$^{(1)}$, Zhi Zhao$^{(1)}$, Manuela Zucknick$^{(1)}$  , Jukka Corander$^{(1),(2)}$}
\address{$\ $\\(1) Oslo Centre for Biostatistics and Epidemiology, Department of Biostatistics, University of Oslo, Norway.
\\
(2) Department of Mathematics and Statistics, University of Helsinki, Finland. \vspace*{.3cm}
\\
Email: t.t.mai@medisin.uio.no
}

\abstract{local low-rank, drug sensitivity, multi-omics, nuclear norm penalization.
\\[17pt]
{\bf Abstract.} The molecular characterization of tumor samples by multiple \textit{omics} data sets of different types or modalities (e.g. gene expression, mutation, CpG methylation) has become an invaluable source of information for assessing the expected performance of individual drugs and their combinations. Merging the relevant information from the omics data modalities provides the statistical basis for determining suitable therapies for specific cancer patients. Different data modalities may each have their own specific structures that need to be taken into account during inference. In this paper, we assume that each \textit{omics} data modality has a low-rank structure with only few relevant features that affect the prediction and we propose to use a composite local nuclear norm penalization for learning drug sensitivity. Numerical results show that the composite low-rank structure can improve the prediction performance compared to using a global low-rank approach or  elastic net regression.
}

\begin{document}
\thispagestyle{myheadings}
\pagestyle{myheadings}
\markright{\tt Proceedings of CIBB 2019}

\section{\bf Introduction}

In recent years, large-scale in-vivo pharmacological profiling of cancer drugs on a panel of cancer cell lines, which are well characterised by multiple \textit{omics} data sets, have been proposed as a promising route to precision medicine ~\cite{Barretina2012, Garnett2012,Ali2018,Yang2013}. The \textit{omics} data can for example consist of genome-wide measurements of mRNA expression, DNA copy numbers, DNA single point and other mutations or CpG methylation of cell lines. These measurements reflect different heterogeneous molecular profiles of the cancer cell lines with respect to driver effects, intra correlations, measurement scales and background noise~\cite{Hasin2017}. The response or sensitivity of a cell line to a drug is characterized by parameters estimated from the dose-response curve, for example by the half maximal inhibitory concentration (IC$_{50}$) \cite{Garnett2012}.

Various supervised machine learning methods have been proposed for the problem of drug response prediction. For example, \cite{Costello2014} utilize a multiple kernel learning (MKL) approach to combine different datasets and explicitly incorporate prior biological information. \cite{Ammad-ud-din2016} combine a MKL approach with matrix factorization, using the model to uncover the latent relationships between drug targets and intracellular pathways. Some other related works can be found in a recent review by \cite{Ali2019machine}.

Combining different data sources and prior biological knowledge can clearly help to shed light on the complexities of cancer drug sensitivity prediction. Most of the previous approaches based on combined multiple \textit{omics} data employ a global structure for parameters inference such as low-rank or sparsity. However, in this application each data source has its own specific structure that is important for capturing the effects of drug sensitivity. Borrowing motivation from the recent work \cite{Li2018inter} in another application domain that explores a composite local low-rank structure, we propose to use a local low-rank model for predicting drug sensitivity with multi-\textit{omics} data. To the best of our knowledge, it is the first time a composite local low-rank structure is applied in the context of drug sensitivity prediction with multi-omics data.

\section{\bf Materials and Methods}

The pharmacological data $ Y=\{y_{ij}\} \in \mathbb{R}^{n\times q} $ represent the sensitivities of $n$ samples (e.g. cell lines or patients) to $q $ drugs. We observe the high-dimensional (multi-omics) data that contain $p_k$ features $ X_k \in  \mathbb{R}^{n\times p_k} $ for $ k = 1, \ldots, K $, and in total all $p = \sum_{k=1}^{K}p_k $ features are available for $ n $ samples across the $K$ different data modalities. Here $\mathbf{X}=  (X_1, \ldots, X_K) \in \mathbb{R}^{n\times p} $ and let's denote the linear model mapping from high-dimensional covariates data to multivariate responses as
\begin{align}
\label{linear.model}
Y = \sum_{k = 1}^{K} X_k B_{k}  + E 
= \mathbf{X} \mathbf{B} + E
\end{align}
where $ \mathbf{B} = ( B^{\top}_{1} , \ldots, B^{\top}_{K})^\top \in \mathbb{R}^{p\times q} $ is the unknown regression coefficient matrix partitioned corresponding to the predictor groups. The random errors $ E $ are assumed to be zero-mean, where specific correlation structures are examined in the simulation study.

Under the model \eqref{linear.model}, we assume each \textit{omics} data set $X_k$ has its own low-rank coefficient matrix $B_k$. Note that this local low-rank assumption does not necessarily imply a low-rank structure of the whole coefficient matrix $\mathbf{B} $.  We propose to use a composite nuclear-norm penalization to estimate a local low-rank structure 
\begin{align}
\label{eq:composite.low.rank}
\hat{B}_{CLR} = \arg\min_{B \in \mathbb{R}^{p\times q} }
\frac{1}{2n} \| Y - \mathbf{X}B \|_2^2 + \lambda \sum_{k=1}^{K} w_k \| B_k \|_*   ,
\end{align}
where $ \lambda >0$ is a tuning parameter and $ w_k $s are the pre-specified weights. Here $ \| A \|_s = (\sum_{ij} |A_{ij}|^s )^{1/s} $ denotes the matrix $\ell_s$-norm and $\| A \|_* = \sum_{j=1}^{{\rm rank} (A)} \sigma_j (A) $ is the nuclear norm with $\sigma_j (\cdot) $ denoting the $ j$th largest singular value of the enclosed matrix. As studied in \cite{Li2018inter}, the weights are used to adjust for the dimension and scale differences of $ X_k $s and the choice 
\begin{align}
\label{weights}
w_k = \sigma_1 (X_k) (\sqrt{q} + \sqrt{ {\rm rank} (X_k)} )/n
\end{align}
balances the penalization of different views and allows us to use only a single tuning parameter.

Remark 1:
Note that problem \eqref{eq:composite.low.rank} covers other well-known problems as its special cases such as:
\begin{itemize}
\item[i)] global low-rank structrure, also known as nuclear norm penalized regression: with $p_1 = p, K = 1$ the penalty in \eqref{eq:composite.low.rank} becomes the nuclear norm penalty of the whole parameter matrix $\mathbf{B}$. 
\item[ii)] multi-task learning: with $p_k = 1, p=K $,  \eqref{eq:composite.low.rank} becomes a special case of multi-task learning that all the tasks share the same set of features and samples.
\end{itemize}

Remark 2: Some theoretical results of the composite local low-rank estimator \eqref{eq:composite.low.rank} have been laid out  in \cite{Li2018inter} for a specific case where $E$ has i.i.d Gaussian entries. More specifically, non-asymptotic bounds for estimation and prediction are given in the Theorem 1 and Theorem 2 in \cite{Li2018inter}.

\section{\bf  Numerical study}
In this section, we conduct simulation studies to examine the efficacy of the proposed composite local low-rank method. The R code to reproduce the results is available at \url{https://github.com/zhizuio/LowRankLearning}.

\subsection{Simulation setups and details}

Using the dimensionalities: $ q = 24, n = 90, K=2, p_1 = p_2 = 100$, the data are simulated as in linear model \eqref{linear.model}, where $ \mathbf{X} = [X_1, X_2]$. Each $X_k (k=1,2)$ is generated from a multivariate normal distribution with mean $\mathbf{0}$ and covariance matrix $\Sigma_X$. The covariance matrix $\Sigma_X $ has the diagonal values equal to 1 and all off-diagonal elements equal to $\rho_X \geq 0$. To take into account the correlation between the drugs, we simulate the noise $E $ from a multivariate normal distribution with mean $\mathbf{0}$ and covariance matrix $\Sigma_\epsilon$. The covariance matrix $\Sigma_\epsilon $ has the diagonal values equal to 1 and all off-diagonal elements equal to $\rho_{\epsilon} \geq 0$. 

We vary different correlation setups in \textit{omics} data $X$ and the drugs as follows:
 \begin{itemize}
 \item[(a).] fix $\rho_X = 0$ and vary $\rho_\epsilon$ as $0.0; 0.3; 0.6$,
 \item[(b).] fix $\rho_\epsilon = 0$ and vary $\rho_X$ as $0.3; 0.6; 0.9$.
 \end{itemize}
Then, for each of the above setups, we consider various settings for the true coefficient matrix $ \mathbf{B} = ( B^{\top}_{1} , B^{\top}_{2}  ) $ as:
\begin{itemize}
\item[S1:] each $\mo{B_{k}, k =1,2}$ is a low-rank matrix with $ {\rm rank}(B_1) = 4,  {\rm rank}(B_2) = 6 $ which is generated as $ B_k = L_{{\rm rank}(B_k)} R^\top_{{\rm rank}(B_k)} $ with the entries of $ L_{{\rm rank}(B_k)} \in \mathbb{R}^{p_k \times {\rm rank}(B_k) } $ and $ R_{{\rm rank}(B_k)} \in \mathbb{R}^{q \times {\rm rank}(B_k) }  $ both generated from $\mathcal{N} (0,1)$.

\item[S2:] $ B_1 $ is low-rank as in S1 and $B_2 $ is a sparse matrix where $50\% $ of the elements are non-zero and simulated from $\mathcal{N} (0,1)$. 

\item[S3:] global low-rank, the whole matrix $ \mathbf{B} $ is a rank-2 matrix  simulated as in S1.
\item[S4:] global sparsity, the whole matrix $ \mathbf{B} $ is sparse where $20\% $ of the elements are non-zero and simulated from $\mathcal{N} (0,1)$. 
\end{itemize}

\begin{table}[!ht]
\footnotesize
\caption{MSPE with fixed $\rho_X = 0$ and $\rho_\epsilon$ is varied. The composite low-rank (CLR) method returns the smallest prediction errors. In general, the prediction errors for the 3 methods are increasing when the correlation between drugs increase.}
\vspace*{-1cm}
\begin{center}
\begin{tabular}{ p{3.99cm} | c c  c | c  c c | c c c}
\hline\hline
  &\multicolumn{3}{ c | }{$\rho_{\epsilon} = 0 $} &   \multicolumn{3}{ c | }{$\rho_{\epsilon} = 0.3 $} & \multicolumn{3}{ c  }{$\rho_{\epsilon} = 0.6 $}  
\\
 		&	CLR	&	GLR  	&		Enet	 
 		&	CLR	&	GLR  	&		Enet	
 		&  CLR	&	GLR  	&		Enet
\\
\hline
setting S1	(local low-rank)	
&	{\bf 0.011}	&	0.517	&	5.111		
& {\bf 0.018}	 &	 0.727	&	 9.172 
& {\bf 0.025} & 1.659 & 9.755
\\
setting S2 (low-rank, sparse)	
&	{\bf 0.011}	&	0.056	&	3.286	
& {\bf 0.018}	 &	 0.066	&	 3.834
& {\bf 0.025} & 0.083 & 4.318

\\
setting S3	 (global low-rank)	
&	{\bf 0.011}	&	0.899	&	12.81	 	
& {\bf 0.018}	 &	 1.228	&	 21.45
&  {\bf 0.025}  & 2.212  & 20.00
\\
setting S4	(global sparsity)	
&	{\bf 0.011}	&	0.023	&	0.237	
& {\bf 0.018}	 &	0.076 	&	 0.321
& {\bf 0.025}	 & 	0.068	&  0.695
\\
\hline\hline
\end{tabular}
\end{center}

\caption{MSEE with fixed $\rho_X = 0$ and $\rho_\epsilon$ is varied. The composite low-rank (CLR) returns the smallest error when there is local low-rank structrure in the data, while the reduced-rank regression  (GLR) returns the smallest error when there is a global structrure. }
\vspace*{-1cm}
\begin{center}
\begin{tabular}{ p{4cm} | c c  c | c  c c  | c c c}
\hline\hline
  &\multicolumn{3}{c | }{$\rho_{\epsilon} = 0 $} &   \multicolumn{3}{ c | }{$\rho_{\epsilon} = 0.3 $} &
  \multicolumn{3}{ c  }{$\rho_{\epsilon} = 0.6 $} 
\\
 		&	CLR	&	GLR  	&		Enet	 
 		&	CLR	&	GLR  	&		Enet	
 		&  CLR	&	GLR  	&		Enet
\\
\hline
setting S1	(local low-rank)		
&	{\bf 1.931}	&	2.753	&	7.788	
& {\bf 1.949}	 &	 2.790	&	 7.271
& {\bf 1.899} & 2.744 & 7.844
\\
setting S2 (low-rank, sparse)	
&	{\bf 0.880} 	&	1.254  	&	3.486	
& {\bf 0.872}	 &	 1.234	&	 3.389
& {\bf 0.895} & 1.278 	& 3.531
\\
setting S3	  (global low-rank)	
&	1.133 &	{\bf 1.121} 	&	3.047	
& 1.173	 &	 {\bf 1.142}	&	 3.061
& 1.177 & {\bf 1.109} 		& 2.868
\\
setting S4	 (global sparsity)	
&	0.121 &	{\bf 0.114}		&	0.319	
& 0.125	 &	 {\bf 0.117}	&	 0.328
& 0.131  & {\bf 0.122}  	&  0.322
\\
\hline\hline
\end{tabular}
\end{center}
\end{table}

We compare the composite local low-rank estimator (\textbf{CLR}) in \eqref{eq:composite.low.rank} with a global low-rank estimator (\textbf{GLR}) for the reduced-rank regression,
\begin{align*}
\hat{B}_{GLR} = \arg\min_{B \in \mathbb{R}^{p\times q} }
\frac{1}{2n} \| Y - \mathbf{X}B \|_2^2, \quad s.t \quad {\rm rank}(B) \leq r
\end{align*}
and the elastic-net (sparsity-inducing) estimator (\textbf{Enet})
\begin{align*}
\hat{B}_{enet} = \arg\min_{B \in \mathbb{R}^{p\times q} }
\frac{1}{2n} \| Y - \mathbf{X}B \|_2^2 + \lambda  (\alpha \| B \|_1 + 0.5 ( 1 - \alpha)  \| B \|^2_2).
\end{align*}
 We use an implementation of the reduced-rank regression from R package 'rrpack'\footnote{\url{https://cran.r-project.org/package=rrpack}} where the rank is chosen by cross-validation. For the elastic-net, we use the R package 'glmnet'\footnote{\url{https://cran.r-project.org/package=glmnet}} with $\lambda $ chosen by 10-folds cross-validation and $\alpha = 0.2 $.

The evaluations are done by using the mean squared estimation error (\textbf{MSEE}) and the mean squared prediction error (\textbf{MSPE})
\begin{align*}
{\rm MSEE} := \frac{1}{pq} \| \hat{B} - \mathbf{B} \|_2^2, 
\quad
{\rm MSPE} := \frac{1}{nq} \| Y - \mathbf{X} \hat{B} \|_2^2 .
\end{align*}
Note that in \mo{real-world applications, where} the true $\mathbf{B} $ is not known, we can only access the prediction errors. We repeat each experiment setting 30 times and report the mean of the outputs.

\subsection{Numerical results}

A first conclusion from the numerical results is that the proposed composite local low-rank (CLR) method has the smallest prediction error. This can be seen from the Table 1 and Table 3. In term of estimation error, the reduced-rank regression (global low-rank method, GLR) seems to be better and also more robust to the correlation between the drugs (Table 2) and between the covariates (Table 4). On the other hand, the Elastic-Net \mo{neither works} well for estimation nor for prediction.

Regarding the prediction error, besides the fact that the CLR method returns the smallest prediction errors, it is also robust to the correlation between the covariates (the \textit{omics} data) as in Table 3. This can be easily seen as  the composite local low-rank method does take into account the local structure of each \textit{omics} data. On the other hand, as the correlation between the drugs increases, the prediction errors of the CLR also increase. Therefore, it would be nice to incorporate the drugs correlation into the preposed prediction method, this idea has been studied in a different model in \cite{ZhaoZucknick2019}.

On the estimation error, the global low-rank (GLR) method works better than CLR and Enet, see Table 2 and Table 4. In particular, the proposed composite local low-rank (CLR) method returns very high estimation errors in all simulation settings (except S4) when the correlation between the covariates increases in Table 4. This could be due to the weights $w_k$ in \eqref{weights} being calculated based on the theoretical study for independent and identical distributed errors, and a restricted eigenvalue condition on $X $, see \cite{Li2018inter}.

\begin{table}[!h]
\footnotesize
\caption{MSPE with fixed $\rho_\epsilon =0 $ and $\rho_X$ is varied. The composite low-rank (CLR) method returns the smallest prediction errors. }
\vspace*{-1cm}
\begin{center}
\begin{tabular}{ p{3.99cm} | c c  c | c  c c | c c c }
\hline\hline
  & \multicolumn{3}{ c | }{$\rho_X = 0.3 $} 
  & \multicolumn{3}{ c | }{$\rho_X = 0.6 $}  
  & \multicolumn{3}{ c  }{$\rho_X = 0.9 $} 
\\
 		&	CLR	&	GLR  	&		Enet	 
 		&	CLR	&	GLR  	&		Enet	
 		&  CLR	&	GLR  	&		Enet
\\
\hline
setting S1	(local low-rank)	
&	{\bf 0.011}	 &	0.524	&	20.40	
& {\bf 0.012}	 &	 0.525	&	 23.01 
& {\bf 0.011} & 0.521 	&  136.5
\\
setting S2 (low-rank, sparse)	
&	{\bf 0.011}	 &	0.041	&		10.10
& {\bf 0.011}	 &	 0.087	&	 12.00
& {\bf 0.011}  & 0.130 	& 30.77

\\
setting S3	 (global low-rank)	
&	{\bf 0.012}	 &	0.895	&	23.34	
& {\bf 0.011}	 &	 0.893	&	 26.29
&  {\bf 0.011}  & 0.891  & 99.65
\\
setting S4	(global sparsity)	
&	{\bf 0.012} &	0.062	&	0.272	
& {\bf 0.011}	 &	0.087 	&	 0.451
& {\bf 0.012}	 & 0.298	&  1.667
\\
\hline\hline
\end{tabular}
\end{center}

\caption{ MSEE with fixed $\rho_\epsilon =0 $ and $\rho_X$ is varied. Overall, the reduced-rank regression (GLR) returns the smallest error. }
\vspace*{-1cm}
\begin{center}
\begin{tabular}{ p{4cm} | c c  c | c  c c  | c c c  }
\hline\hline
  & \multicolumn{3}{ c | }{$\rho_{X} = 0.3 $} &
  \multicolumn{3}{ c | }{$\rho_{X} = 0.6 $} &
  \multicolumn{3}{ c  }{$\rho_{X} = 0.9 $} 
\\
 		&	CLR	&	GLR  	&		Enet	 
 		&	CLR	&	GLR  	&		Enet	
 		&  CLR	&	GLR  	&		Enet
\\
\hline
setting S1	(local low-rank)		
&	{\bf 2.165}	&	2.755	&	7.640	
& 4.691	 &	 {\bf 2.752}	&	 7.855
& 14.60 & {\bf 2.726} & 7.692
\\
setting S2 (low-rank, sparse)	
&	1.153 &	{\bf 1.298}  	&	3.581	
& 1.828	 &	 {\bf 1.243}	&	 3.507
& 5.447 & {\bf 1.186} & 3.266
\\
setting S3	  (global low-rank)	
&	1.213 	&	{\bf 1.131} 	&	3.014	
& 1.235  &	 {\bf 1.054}	&	 2.860
& 4.225 & {\bf 1.137} 	& 3.167
\\
setting S4	 (global sparsity)	
&	0.121 &	{\bf 0.112}		&	0.331	
& 0.120	 &	 {\bf 0.112}	&	0.331
& 0.119  & {\bf 0.112}  &  0.335
\\
\hline\hline
\end{tabular}
\end{center}
\end{table}

\section{\bf Real data analysis: GDSC data}

To test our approach on a real dataset, we use data from a large-scale pharmacogenomic study, the Genomics of Drug Sensitivity in Cancer (CDSC) \cite{Yang2013}, made available online\footnote{\url{ftp://ftp.sanger.ac.uk/pub4/cancerrxgene/releases/release-5.0/}} by Garnett \textit{et al}. \cite{Garnett2012}. The dataset consists of drug sensitivity data (IC$_{50}$) for a large panel of cancer drugs screened on multiple cell lines, in addition to various omics measurements from the cancer cell lines.

We select a subset of the data, consisting of 97 screened drugs and 498 human cancer cell lines from 13 cancer tissue types, such that the response matrix $Y \in \mathbb{R} ^{498\times 97}$ is fully observed. For each cell line, they measured the mutation \mo{status for a panel of known cancer genes, and genome-wide} copy number variation and gene expression. \mo{For data preprocessing and feature selection, we follow the procedure in \cite{ZhaoZucknick2019}, which results in preselecting 68 binary mutation features ($X_1 \in \mathbb{R}^{498 \times 68} $), 426 integer copy number features ($X_2 \in \mathbb{R}^{498 \times 426} $) and 2602 continuous gene expression features ($X_3 \in \mathbb{R}^{498 \times 2602} $), respectively, and the drug sensitivity being measured as $\log\mbox{IC}_{50}$}. Note that, we did not consider 13 cancer tissue indicators in our analysis as it is non-omics data. 

The performances of 3 methods on the GDSC data set are given in Table 5. The estimated ranks of each omics data source $B_1 \in \mathbb{R}^{2602\times 97}, B_2 \in \mathbb{R}^{426 \times 97} , B_2 \in \mathbb{R}^{68 \times 97}  $ from the composite local low-rank model are 97, 54 and 61 respectively. For reduced-rank regression (GLR), we fit model with maximum rank to be 50 and the best selected rank is 50. GLR returns smallest prediction error as, from the simulation, it is more robust to the correlation among the drugs. For elastic net, 99\% of the coefficient is estimated to be zero. More specifically, among the smallest predictions drugs, CLR reports some drugs with smaller prediction errors comparing to GLR: for example prediction error of Bicalutamide drug by CLR is 0.003 while GLR returns 0.046.

\begin{table}[!ht]
\footnotesize
\caption{MSPE with real data.}
\begin{center}
\begin{tabular}{ p{2cm} | c c  c  }
 		&	CLR	&	GLR (rank-50)  	&		Enet	 
\\
\hline
real data	&	 0.3340 	&	0.1257   &  0.7475
\end{tabular}
\end{center}
\end{table}

\section{\bf Discussion and Conclusion}
In this paper, we have studied the problem of drug sensitivity prediction with multi-omics data. Under the assumption that each \textit{omics} data modality affects the drug sensitivity prediction only through a few latent features (low-rankness), we propose to use a composite local low-rank model that takes into account this local structure. Our numerical results illustrate beneficial performance regarding the prediction errors of the proposed method compared to global methods, such as reduced-rank regression and elastic net.

This paper represents an initial take on the drug prediction based on local low-rank structures. There are some clear limitations in our approach, such as: (i) incorporating correlations between drugs and the heterogeneity of multi-omics data, as in \cite{ZhaoZucknick2019}, into our model would help to make our method more robust; (ii) incorporating other local structure rather than low-rankness, could help our method to become more flexible; (iii) extending the proposed method by including full-rank ``mandatory'' non-omics data sources  (e.g clinical variables). These problems open further venues of research in this area in the future.

\section*{\bf Acknowledgments}

The first two authors contributed equally. L.R. is supported by The Norwegian Research Council 237718 through the Big Insight Center for research-driven innovation. The research of T.T.M. and J.C. are supported by the European Research Council (SCARABEE, no. 742158).

\bibliographystyle{apalike}
{\fontsize{10}{10}\selectfont

}
\end{document}